\documentclass{iopart}
\usepackage{graphicx} %eps figures can be used instead
\usepackage{color} %eps figures can be used instead
\newcommand{\ib}[1]{{\color{black}#1}}

\newcommand{\alt}{\raisebox{-0.3ex}{$\stackrel{<}{\sim}$}}
\newcommand{\agt}{\raisebox{-0.3ex}{$\stackrel{>}{\sim}$}}

\begin{document}

% \title[A short title]{Quantifying the molecular orbital alignment relative to electrodes Fermi energy for molecular junctions with similar chemical linkage}
\title[Relative orbital alignment for molecular junctions with similar chemical linkage]{Quantifying the relative molecular orbital alignment for molecular junctions with similar chemical linkage to electrodes}

\author{Ioan B\^aldea}

\address{Theoretische Chemie, Universit\"at Heidelberg, Im Neuenheimer Feld 229, D-69210 Heidelberg, Germany
}
\ead{ioan@pci.uni-heidelberg.de. Also at Institute of Space Sciences, National Institute for Lasers, Plasmas and Radiation Physics, Bucharest, Romania}
\begin{abstract}
Estimating the relative alignment between the frontier molecular orbitals
that dominates the charge transport through single-molecule junctions represents
a challenge for theory. This requires approaches beyond the widely employed framework provided by 
the density functional theory, wherein the Kohn-Sham ``orbitals'' are treated as if they were
real molecular orbitals, which is not the case. In this paper, we report results obtained 
by means of quantum chemical calculations, including the 
EOM-CCSD (equation-of-motion coupled-cluster singles and doubles), 
which is the state-of-the-art of quantum chemistry for medium-size molecules like 
those considered here. These theoretical results are validated
against data on the molecular orbital energy offset relative to the electrodes' Fermi energy 
extracted from experiments for junctions based on
4,4'-bipyridine and 1,4-dicyanobenzene.
\end{abstract}

% \vspace{2pc}
\maketitle

\section{Introduction}
The last decades marked significant advances in the fabrication and characterization of a variety of 
molecular electronic devices based 
on a single or a small number of molecules. For understanding an impressive amount of experimental 
material accumulated, results of numerical calculations are usually presented.
They are performed within theoretical frameworks, which are often completely opaque and preclude  
a straightforward interpretation in terms of properties having a simple and clear physical meaning.

An important property of this kind is the relative alignment of the frontier 
orbitals relative to the electrodes' Fermi energy 
$E_F$ of a molecule embedded in a molecular junction \cite{Datta:03}.
The energy offset $\varepsilon_0 = \min\left(E_{LUMO} - E_F, E_F - E_{HOMO}\right)$ 
of the highest occupied or lowest unoccupied molecular orbital (HOMO or LUMO, respectively),
whichever is closest to the Fermi level $E_F$, is usually compared to a 
tunneling energy barrier \cite{Reed:09}. It is a key quantity in molecular transport, 
because it controls the charge transfer efficiency. Current experimental methods 
to estimate the energy offset $\varepsilon_0$ employ ultraviolet photoelectron spectroscopy (UPS)
 \cite{Frisbie:11}, thermopower \cite{Datta:03b,Reddy:08,Venkataraman:12a,Tao:13},
and transition voltage spectroscopy (TVS) \cite{Beebe:06}.
TVS is a method that became very popular in the molecular electronic community
due to its simplicity \cite{Beebe:08,Reed:09,Hacker:09,Reed:11,Guo:11,Lee:11,Reddy:11,Lennartz:11,Vuillaume:12c,Fracasso:13}.
TVS-based results deduced from ref.~\cite{Reed:09}, a work that significantly 
contributed to the TVS popularity, represent  
an essential piece of experimental data to be used in the present study.

Postulating the existence of a single orbital that dominates the charge transport in 
a molecular junction might appear to be a too crude approximation.
However, recent extensive analysis 
\cite{Baldea:2010h,Baldea:2012a,Baldea:2012b,Baldea:2012g,Baldea:2013b,Baldea:2013d,Baldea:2014a}
of existing transport data measured for a variety of molecular junctions demonstrated that, 
in the entire voltage accessed experimentally, current-voltage ($I-V$) curves can indeed be excellently be reproduced 
by assuming just the contribution of a single dominant molecular orbital (MO). 
Although this is an enormous simplification,
the quantitative description of the relative alignment
$\varepsilon_0$ of the dominant orbital of the embedded molecule remains an important challenge for \emph{ab initio}
approaches to the charge transport. The vast majority of the theoretical approaches of the charge transport  
through molecular devices utilized to date are based on the combination of the 
nonequilibrium Green's functions (NEGF) and density functional theory (DFT) \cite{Xue:02}. 
The most important drawback of
such approaches directly related to the main issue considered in this paper is the fact that they treat 
the eigenvalues of the Kohn-Sham equations as if they were energies with physical meaning.
In reality, as is well known, Kohn-Sham orbitals are mathematical objects rather than 
true molecular orbitals \cite{Gunnarson:89}.
So, it is not at all surprising that their ``energies'' 
cannot provide an adequate physical description
\cite{Gunnarson:89,Baldea:2014c}. Unoccupied orbitals are especially difficult to describe 
theoretically \cite{Baldea:2013b,Baldea:2014a}. Therefore, 
molecular junctions exhibiting an n-type (LUMO-mediated) conduction, 
like the ones to be considered in this study, 
deserve a special consideration. 

Demonstrating that valuable information on the LUMO alignment  
in molecular junctions can be obtained within reliable quantum chemical approaches beyond DFT
represents an important aim of the present paper. To validate the theoretical approach developed here,
we will employ experimental data for molecular junctions based on 
1,4-dicyanobenzene (BDCN) \cite{Reed:09} and 4,4'-bipyridine (44BPY) \cite{Venkataraman:12a}.
%%%%%%%%%%%%%%%%%%%%%%%%%%%%%%%%%%%%%%%%%%%%%%%%%%%%%%%%%%%%%%%%%%%%%%%%%%%%%%%%%%%%%%%%%
\section{Disentangling the contributions to the dominant MO energy offset}
\label{sec:disentangling}
%%%%%%%%%%%%%%%%%%%%%%%%%%%%%%%%%%%%%%%%%%%%%%%%%%%%%%%%%%%%%%%%%%%%%%%%%%%%%%%%%%%%%%%%%
As a remedy of the main drawback of approaches to the charge transport 
through molecular devices that combine nonequilibrium Green's functions (NEGF) and DFT mentioned in
Introduction, 
in more elaborate (so-called ``DFT+$\Sigma$'') developments \cite{Quek:07,Venkataraman:12a}
the occupied 
($\epsilon_p < E_F$) and unoccupied ($\epsilon_p > E_F$) 
KS ``orbitals'' are rigidly shifted in opposite directions by the same 
($p$-independent) amount $\Delta$ obtained 
either by fitting experimental data \cite{Ferretti:11} 
or in a two-step procedure 
as follows. First, a value $\Delta_0 = E_{LUMO} - E_{HOMO}$ 
of the HOMO-LUMO gap is deduced from the energies $\mathcal{E}$
of the various (neutral, anionic and cationic) charge species in the gas phase
$\Delta_0 = \mathcal{E}_{anion} + \mathcal{E}_{cation} - 2 \mathcal{E}_{neutral}$,
a method known as $\Delta$-SCF \cite{Gunnarson:89} (or, more appropriate, $\Delta$-DFT 
\cite{Baldea:2012i,Baldea:2013a,Baldea:2013b,Baldea:2014c}). 
Because the value $\Delta_0$ is usually
much too large, it is then renormalized ($\Delta_0 \to \Delta$) by considering 
image charges of a LUMO (HOMO) modeled as a point-like electron (hole)
formed in electrodes taken as infinite plates in the immediate vicinity of the active molecule 
\cite{Neaton:06,Quek:07,Venkataraman:12a}. 
Although the renormalization found in this way may render 
the corresponding $\Delta$-value compatible to experiments,
recent work \cite{Baldea:2012e,Baldea:2012f,Baldea:2013c,Baldea:2014c} has drawn 
attention on the fact that these assumptions are inadequate
for realistic molecular junctions \cite{Baldea:2012e,Baldea:2012f,Baldea:2013c,Baldea:2014c}.

An aspect on which we want to draw attention in this study is the following. 
NEGF-DFT transport calculations done as described above utilize an extended molecule, which 
includes several atomic layers from electrodes in addition to the active molecule. 
On the other side, as it has been long recognized, 
the classical image potential originates from the interaction of the electron in the LUMO with 
electronic collective (long-wavelength polarization) 
modes in the metals, in particular, surface, interface and bulk plasmons 
\cite{Ritchie:72,Inkson:73,Lenac:76,Jonson:80,Sunjic:91,Ness:98a}.
So, the effect of the electrons of the atoms of the electrodes belonging to the extended molecule
is also accounted for in the interaction energy $\Phi_{im}$ with the image charges.
Therefore, the procedure of applying $\Sigma$-corrections (these corrections are due to image charges) 
on top of NEGF-DFT approaches is plagued by double counting. 

To overcome this drawback, we propose here a disentangling procedure, which we then validate 
by comparing molecular junctions based on two molecules, 
namely 1,4-dicyanobenzene (BDCN) and 4,4'-bipyridine (44BPY). 
Previous studies demonstrated an n-type (LUMO-mediated) conduction for both types of junctions 
\cite{Reed:09,Tao:03,Venkataraman:12a,Baldea:2013b}. So, it is the LUMO on which attention will be focused below.

%%%%%%%%%%%%%%%%%%%%%%%%%%%%%%%%%%%%%%%%%%%%%%%%%%%%%%%%%%%%%%%%%%%%%%%%%%%%%%%% 
\begin{figure}[h!]
$ $\\[0ex]
\centerline{\hspace*{0.ex}\includegraphics[width=0.5\textwidth,angle=0]{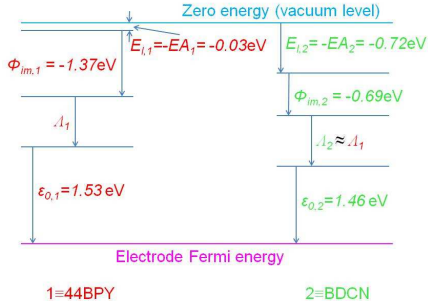}\hspace*{0.ex}}
$ $\\[-1.4ex]
\caption{Disentangling the contributions to the LUMO energy offset for the two molecules 
--- 4,4'-bipyridine (44BPY) and 1,4-dicyanobenzene (BDCN) --- 
embedded in the molecular junctions considered in the present paper. 
The electron attachment energy with reversed sign represents the LUMO energy 
of the isolated molecule ($E_{l} = -EA$). 
$\Phi_{im}$ and $\Lambda$ represent LUMO energy shifts due to image charges and local 
(contact) effects resulting from long-range and short-range interactions between the embedded molecule and
electrodes, respectively. The nearly equal quantities  $\Lambda_1 \approx \Lambda_2$ 
reflect the similar chemical linkage (nitrogen-gold affinity) to electrodes.
See the main text for details.}
\label{fig:lumo-energies}
\end{figure}
%%%%%%%%%%%%%%%%%%%%%%%%%%%%%%%%%%%%%%%%%%%%%%%%%%%%%%%%%%%%%%%%%%%%%%%%%%%%%%%%

The disentangling scheme for the LUMO energy is presented in Figure \ref{fig:lumo-energies}. 
(For the numerical values given in Figure \ref{fig:lumo-energies} please refer to Sec.~\ref{sec:results}.)
In the absence of molecule-electrode couplings, the LUMO energy $E_l$ is given by the lowest 
electron attachment energy $EA$ of the isolated molecule taken with reversed sign 
($E_l = -EA$, Koopman's theorem). The LUMO energy $E_{LUMO}$ of the 
molecule embedded in a nanojunction differs from that of the isolated molecule
because an electron transferred to the LUMO interacts with the electrodes. The scheme proposed here
holds in cases (and we will show below that such cases do exist) where it is possible to
split this interaction into a short-range and a long-range part that can be analyzed separately. 
They yield two contributions
(denoted $\Lambda$ and $\Phi_{im}$, respectively) to the corresponding LUMO energy shift
%%%%%%%%%%%%%%%%%%%%%%%%%%%%%%%%%%%%%%%%%%%%%%%%%%%%%%%%%%%%%%%%%%%%%%%%%%%%%%%% 
\begin{equation}
\label{eq-lumo-shift}
E_{LUMO} = E_{l} + \Lambda + \Phi_{im} .
\end{equation}
%%%%%%%%%%%%%%%%%%%%%%%%%%%%%%%%%%%%%%%%%%%%%%%%%%%%%%%%%%%%%%%%%%%%%%%%%%%%%%%% 
Eq.~(\ref{eq-lumo-shift}) allows one to express the LUMO energy offset as
%%%%%%%%%%%%%%%%%%%%%%%%%%%%%%%%%%%%%%%%%%%%%%%%%%%%%%%%%%%%%%%%%%%%%%%%%%%%%%%% 
\begin{equation}
\label{eq-lumo-offset}
\varepsilon_{0} = E_{LUMO} - E_F = -EA + \Lambda + \Phi_{im} - E_F .
\end{equation}
%%%%%%%%%%%%%%%%%%%%%%%%%%%%%%%%%%%%%%%%%%%%%%%%%%%%%%%%%%%%%%%%%%%%%%%%%%%%%%%% 

To avoid the double counting issue mentioned above, 
in addition to the electrodes' collective effect embodied in the image charge contribution $\Phi_{im}$,
we will consider the \emph{local} LUMO energy shift $\Lambda$ due to the interaction of the active molecule
with the metal atoms at the its ends. This accounts for the well known fact that a 
chemisorbed molecule strongly coupled to
the substrate often has valence molecular orbital energies substantially different 
from that in the gas phase \cite{Gustaffson:78}. 
Experimental data 
on molecular junctions, indicating a substantial MO energy shift due to local
(interface) dipoles \cite{Frisbie:11}, may also be taken as a confirmation of this hypothesis.
%%%%%%%%%%%%%%%%%%%%%%%%%%%%%%%%%%%%%%%%%%%%%%%%%%%%%%%%%%%%%%%%%%%%%%%%%%%%%%%%%%%%%%%%%
\section{Method}
\label{sec:method}
%%%%%%%%%%%%%%%%%%%%%%%%%%%%%%%%%%%%%%%%%%%%%%%%%%%%%%%%%%%%%%%%%%%%%%%%%%%%%%%%%%%%%%%%%
To obtain the theoretical results 
% for the electron affinities ($EA$) 
presented in this paper we have utilized the 
EOM-CCSD (equation-of-motion coupled-cluster singles and doubles) 
\cite{Stanton:95,Nooijen:95a,Nooijen:95b}. 
This method represents the state-of-the-art of quantum chemistry for medium-size 
molecules, like the ones to be considered here. The CCSD calculations  
% total energies of the various charge species and excitation energies 
were performed with the CFOUR package \cite{cfour}. 
For comparison purposes, results based on hybrid coupled clusters (CC2) \cite{Christiansen:95}
and regular (strict) second-order algebraic-diagrammatic constructions [ADC(2)] \cite{Schirmer:82,Schirmer:91}
will also be presented.
ADC(2) calculations have been done with the fully parallelized PRICD-$\Sigma$(2) code
\cite{Vysotskiy:10a}, which is interfaced to MOLCAS \cite{molcas}. 

The molecular geometries were
optimized at the DFT level using the B3LYP hybrid functional as implemented in GAUSSIAN 09 \cite{g09},
a package also employed to estimate the electroaffinities by means of $\Delta$-DFT calculations
\cite{Gunnarson:89,Baldea:2012i,Baldea:2014c}. All results of the quantum chemical calculations 
reported here were obtained by employing 
aug-cc-pVDZ (Dunning augmented correlation consistent double zeta) basis sets.
As shown in recent studies \cite{Baldea:2013a,Baldea:2013b,Klaiman:13,Baldea:2014c}, these
basis sets include sufficient diffuse basis functions to properly describe anionic states,
and the corresponding results for electron attachment energies can be trusted. 
%%%%%%%%%%%%%%%%%%%%%%%%%%%%%%%%%%%%%%%%%%%%%%%%%%%%%%%%%%%%%%%%%%%%%%%%%%%%%%%%
\section{Results and discussion}
\label{sec:results}
%%%%%%%%%%%%%%%%%%%%%%%%%%%%%%%%%%%%%%%%%%%%%%%%%%%%%%%%%%%%%%%%%%%%%%%%%%%%%%%%%%%%%%%%%
Obtaining estimates for $\Lambda$ from quantum chemical calculations by comparing 
the LUMO energies of an isolated molecule and the same molecule with one 
or a few metal atoms attached at each of its ends will be the object of a further investigation.
Here we will confine ourselves to quantify the difference in the LUMO energy offsets 
$\varepsilon_{0,2} - \varepsilon_{0,1}$ for two molecular junctions consisting of molecules
(labeled $1$ and $2$) joined to electrodes of identical metals ($E_{F,1} = E_{F, 2} = E_F$)
by \emph{similar} chemical linkage (nitrogen-gold affinity in the specific situations discussed below). 
In this case
%%%%%%%%%%%%%%%%%%%%%%%%%%%%%%%%%%%%%%%%%%%%%%%%%%%%%%%%%%%%%%%%%%%%%%%%%%%%%%%% 
\begin{equation}
\label{eq-Lambdas}
\Lambda_{1} \approx \Lambda_{2} ,
\end{equation}
%%%%%%%%%%%%%%%%%%%%%%%%%%%%%%%%%%%%%%%%%%%%%%%%%%%%%%%%%%%%%%%%%%%%%%%%%%%%%%%% 
and Eq.~(\ref{eq-lumo-offset}) yields 
%%%%%%%%%%%%%%%%%%%%%%%%%%%%%%%%%%%%%%%%%%%%%%%%%%%%%%%%%%%%%%%%%%%%%%%%%%%%%%%% 
\begin{equation}
\label{eq-key}
\underbrace{\varepsilon_{0,2} - \varepsilon_{0,1}}_{\mbox{from exp.}} \approx 
\underbrace{EA_{1} - EA_{2} + \Phi_{im,2} - \Phi_{im, 1}}_{\mbox{from theory}} .
\end{equation}
%%%%%%%%%%%%%%%%%%%%%%%%%%%%%%%%%%%%%%%%%%%%%%%%%%%%%%%%%%%%%%%%%%%%%%%%%%%%%%%%
Here, the under braces indicate the method to be used below for evaluating the
corresponding quantities; the LHS can be estimated from available experimental data, 
the RHS can be obtained theoretically via quantum chemical calculations.

To validate the disentangling scheme proposed here, on which the basic Eq.~(\ref{eq-key}) 
relies, we will consider molecular junctions based on 
BDCN and 44BPY.
%%%%%%%%%%%%%%%%%%%%%%%%%%%%%%%%%%%%%%%%%%%%%%%%%%%%%%%%%%%%%%%%%%%%%%%%%%%%%%%%
\subsection{Quantities estimated from experimental data}
\label{sec:exp}
%%%%%%%%%%%%%%%%%%%%%%%%%%%%%%%%%%%%%%%%%%%%%%%%%%%%%%%%%%%%%%%%%%%%%%%%%%%%%%%%%%%%%%%%%
The LUMO energy offset for the BDCN molecule can be deduced from the experimental 
value of the transition voltage $ V_{t} \to V_{t,1} = 1.69  \pm 0.05$\,V extracted from 
the current-voltage ($I-V$) curve measured at zero gate potential ($V_G = 0$); see the supplementary 
information of ref.~\cite{Reed:09}. Let us briefly remind that the transition voltage 
$V_t$ represents the bias at the minimum of the Fowler-Nordheim quantity 
$\ln (I/V^2)$ \cite{Beebe:06}. Because the experimental $I-V$ curve \cite{Reed:09}
turned out to be practically symmetric [$I(V) \simeq - I(-V)$], the LUMO energy offset 
can be estimated as \cite{Baldea:2012a}
%%%%%%%%%%%%%%%%%%%%%%%%%%%%%%%%%%%%%%%%%%%%%%%%%%%%%%%%%%%%%%%%%%%%%%%%%%%%%%%% 
\begin{equation}
\label{eq-vt}
\varepsilon_{0} = \frac{\sqrt{3}}{2} V_{t} ,
\end{equation}
%%%%%%%%%%%%%%%%%%%%%%%%%%%%%%%%%%%%%%%%%%%%%%%%%%%%%%%%%%%%%%%%%%%%%%%%%%%%%%%%
which yields \cite{slight-difference} 
%%%%%%%%%%%%%%%%%%%%%%%%%%%%%%%%%%%%%%%%%%%%%%%%%%%%%%%%%%%%%%%%%%%%%%%%%%%%%%%% 
\begin{equation}
\label{eq-e0-1}
\varepsilon_{0, 1} = 1.46 \pm 0.04\mbox{\,eV} . 
\end{equation}
%%%%%%%%%%%%%%%%%%%%%%%%%%%%%%%%%%%%%%%%%%%%%%%%%%%%%%%%%%%%%%%%%%%%%%%%%%%%%%%%

On the other side, the LUMO energy offset for 44BPY-based junctions deduced via 
thermopower data \cite{Venkataraman:12a,Baldea:2013b} is 
%%%%%%%%%%%%%%%%%%%%%%%%%%%%%%%%%%%%%%%%%%%%%%%%%%%%%%%%%%%%%%%%%%%%%%%%%%%%%%%% 
\begin{equation}
\label{eq-e0-2}
\varepsilon_{0, 2} = 1.53 \pm 0.08\mbox{\,eV} . 
\end{equation}
%%%%%%%%%%%%%%%%%%%%%%%%%%%%%%%%%%%%%%%%%%%%%%%%%%%%%%%%%%%%%%%%%%%%%%%%%%%%%%%%
%%%%%%%%%%%%%%%%%%%%%%%%%%%%%%%%%%%%%%%%%%%%%%%%%%%%%%%%%%%%%%%%%%%%%%%%%%%%%%%%
\subsection{Quantities estimated via quantum chemical calculations}
\label{sec:theory}
%%%%%%%%%%%%%%%%%%%%%%%%%%%%%%%%%%%%%%%%%%%%%%%%%%%%%%%%%%%%%%%%%%%%%%%%%%%%%%%%%%%%%%%%%
%%%%%%%%%%%%%%%%%%%%%%%%%%%%%%%%%%%%%%%%%%%%%%%%%%%%%%%%%%%%%%%%%%%%%%%%%%%%%%%%
\subsubsection{Electron attachment energies}
\label{sec:qc}
%%%%%%%%%%%%%%%%%%%%%%%%%%%%%%%%%%%%%%%%%%%%%%%%%%%%%%%%%%%%%%%%%%%%%%%%%%%%%%%%%%%%%%%%%
The results of the quantum chemical calculations 
for the lowest electron attachment energies $EA_{1,2}$ entering Eq.~(\ref{eq-key})
are collected in Table \ref{table}. 
%%%%%%%%%%%%%%%%%%%%%%%%%%%%%%%%%%%%%%%%%%%%%%%%%%%%%%%%%%%%%%%%%%%%%%%%%%%%%%%%%%%%%%%%%
\begin{table*}[h!]
\begin{center}
\begin{tabular*}{0.9\textwidth}{@{\extracolsep{\fill}}lccc}
\hline
Method   & EA$_{BDCN}$ (eV) &  EA$_{44BPY}$ (eV) &  EA$_{BDCN}$ -  EA$_{44BPY}$ (eV)\\
EOM-CCSD      &  0.717     &  0.032             & 0.685 \\
EOM-CC2       &  1.047     &  0.360             & 0.687 \\
ADC(2)        &  1.107     &  0.370             & 0.737 \\
$\Delta$-CCSD &  0.678     &  0.0043            & 0.678 \\
$\Delta$-DFT  &  1.127     &  0.444             & 0.683 \\
% $\Delta$-MP2  &  0.394     & -0.500             & 0.894 \\
\hline
\hline
\end{tabular*}
\caption{Results for the electron attachment energies $EA$ of the isolated molecules (BDCN and 44BPY) 
considered in this study computed by means of various quantum chemical calculations 
indicated in the left column and described in the main text. The geometries of the neutral molecules have been 
optimized at DFT/B3LYP/aug-cc-pVDZ level. Notice that although the absolute values $EA_{BDCN}$ and $EA_{44BPY}$
may significantly depend on the method utilized, their difference (last column) is quite insensitive to it.
}
\label{table}
\end{center}
\end{table*}
%%%%%%%%%%%%%%%%%%%%%%%%%%%%%%%%%%%%%%%%%%%%%%%%%%%%%%%%%%%%%%%%%%%%%%%%%%%%%%%%%%%%%%
In addition to the values obtained within the 
EOM-CCSD, EOM-CC2, and ADC(2) methods described in Sec.~\ref{sec:method},
values obtained via energy difference ($\Delta$-) CCSD and DFT methods \cite{Baldea:2014c} 
are also presented there. In the latter, the lowest electron attachment energy
is estimated as the difference between the ground state energies ($ \mathcal{E}$)
of the neutral and anionic species at the equilibrium geometry of the neutral molecule
($M$=CCSD, DFT)
%%%%%%%%%%%%%%%%%%%%%%%%%%%%%%%%%%%%%%%%%%%%%%%%%%%%%%%%%%%%%%%%%%%%%%%%%%%%%%%%
\begin{equation}
\mbox{EA}_{M} = \mathcal{E}_{M, neutral} - \mathcal{E}_{M, anion} .
\label{eq-delta-a}
\end{equation}
%%%%%%%%%%%%%%%%%%%%%%%%%%%%%%%%%%%%%%%%%%%%%%%%%%%%%%%%%%%%%%%%%%%%%%%%%%%%%%%%
The inspection the values given in Table \ref{table} is instructive. It reveals that
although the absolute values of the electroaffinities 
$EA_{1} \equiv EA_{BDCN}$ and $EA_{2} \equiv EA_{44BPY}$ for the 
two molecules considered significantly depend on the quantum chemical method
utilized, the differences $EA_{1} - EA_{2}$ deduced by using the aforementioned methods 
is within the experimental accuracy
(\emph{cf.~} Eqs.~(\ref{eq-e0-1}) and (\ref{eq-e0-2})). 
%%%%%%%%%%%%%%%%%%%%%%%%%%%%%%%%%%%%%%%%%%%%%%%%%%%%%%%%%%%%%%%%%%%%%%%%%%%%%%%%
\subsubsection{Image charge effects}
\label{sec:images}
%%%%%%%%%%%%%%%%%%%%%%%%%%%%%%%%%%%%%%%%%%%%%%%%%%%%%%%%%%%%%%%%%%%%%%%%%%%%%%%%%%%%%%%%%
An extensive analysis of the effect of image charges in a two terminal setup
was presented recently \cite{Baldea:2013b}. Therefore, only a few details will be given here.
The interaction energy between two infinite planar electrodes and a point charge $e$ located at $z$
\emph{in vacuo} can be expressed as
%%%%%%%%%%%%%%%%%%%%%%%%%%%%%%%%%%%%%%%%%%%%%%%%%%%%%%%%%%%%%%%%%%%%%%%%%%%%%%
\begin{eqnarray}
\phi_{im}(z) & = &
\frac{e^2}{4 d} \left\{ - 2 \psi(1)
+ \psi\left(\frac{z - z_s}{d}\right)
\left[1 - e^{ - \mu\left( z - z_s\right)}\right] \right . \nonumber \\
& & + \left . \psi\left(\frac{z_t - z}{d}\right)
\left[1 - e^{ - \mu\left( z_t - z\right)}\right] \right\} ,
\label{eq-phi-i}
\end{eqnarray}
%%%%%%%%%%%%%%%%%%%%%%%%%%%%%%%%%%%%%%%%%%%%%%%%%%%%%%%%%%%%%%%%%%%%%%%%%%%%%%
where $\psi(z) \equiv d\,\ln \Gamma(z)/d\,z$ is the digamma function.
Eq.~(\ref{eq-phi-i}) is obtained from the expression 
deduced within classical electrostatics \cite{Sommerfeld:33}
by inserting the expressions in 
the square parentheses, which ensure that the limits $\lim_{z \to z_{s, t}}\phi_{im}(z)$ remain finite
and provide good fits of the microscopically calculated
potential for the single-plane problem ($z \agt z_s$, $z \alt z_t$)
\cite{JJJ:84,Smith:89}. The positions $z_{s,t}$ ($z_{s} < z_{t}$) of the image planes
are outwardly shifted by $z_0$ from the electrode surfaces $z_{s,t}^{\prime} = z_{s,t} \mp z_0$
\cite{Lang:73,desjonqueres:96}, where $z_0$ represents a quantum correction to the classical result.
Numerical values appropriate for gold electrodes [Au(111) faces] are 
$\mu\simeq 1.25$\,bohr$^{-1}$ and $z_0 \simeq 1.58$\,{\AA} \cite{Baldea:2013b}.

\begin{figure}[h!]
$ $\\[0ex]
\centerline{\hspace*{0.ex}\includegraphics[width=0.5\textwidth,angle=0]{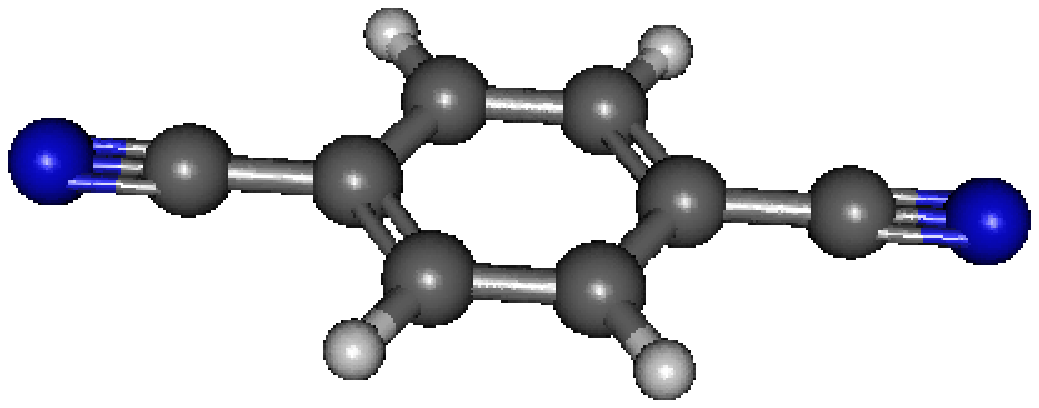}\hspace*{0.ex}\hspace*{0.ex}\includegraphics[width=0.5\textwidth,angle=0]{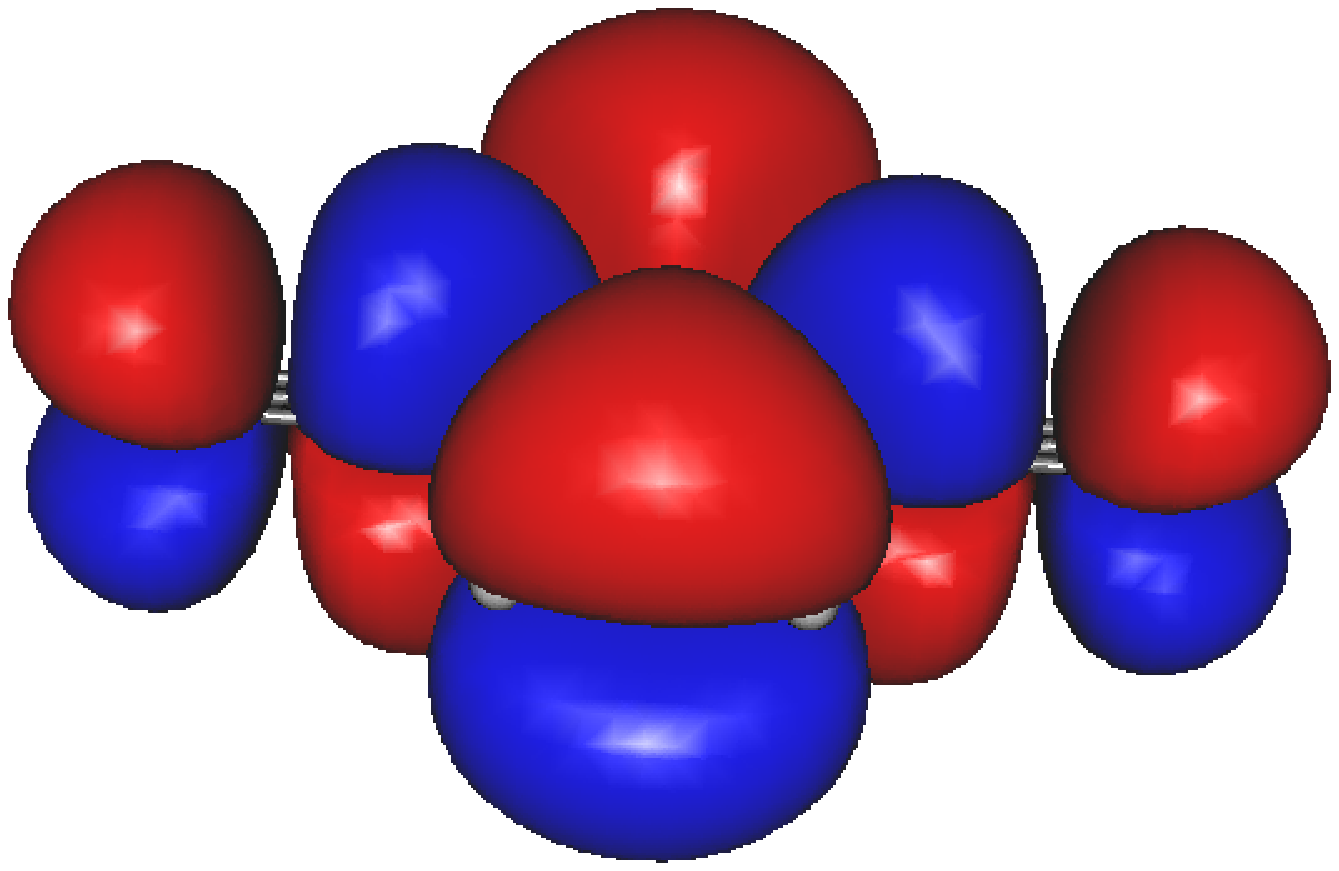}\hspace*{0.ex}}
$ $\\[-1.4ex]
\caption{The almost singly occupied natural orbital corresponding to the anion's extra electron 
of the BDCN$^{\bullet -}$ anion (``LUMO'')
obtained via EA-EOM-CCSD/aug-cc-pVDZ calculations is delocalized over the whole molecule.
The LUMO density presented here and in Figure \ref{fig:lumo-44bpy} was generated by using Gabedit \cite{gabedit}.}
\label{fig:lumo-bdcn}
\end{figure}

\begin{figure}[h!]
$ $\\[0ex]
\centerline{\hspace*{0.ex}\includegraphics[width=0.5\textwidth,angle=0]{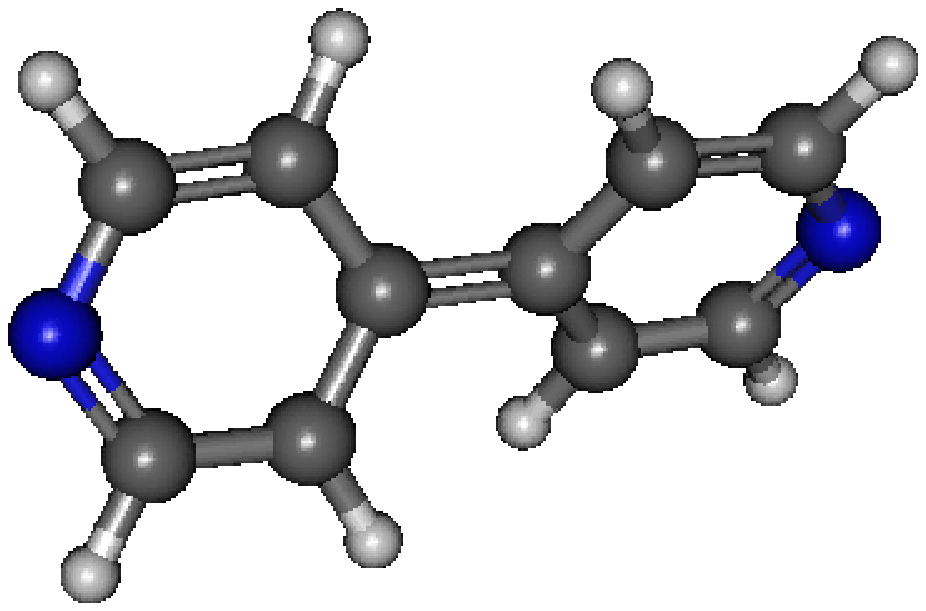}\hspace*{0.ex}\hspace*{0.ex}\includegraphics[width=0.5\textwidth,angle=0]{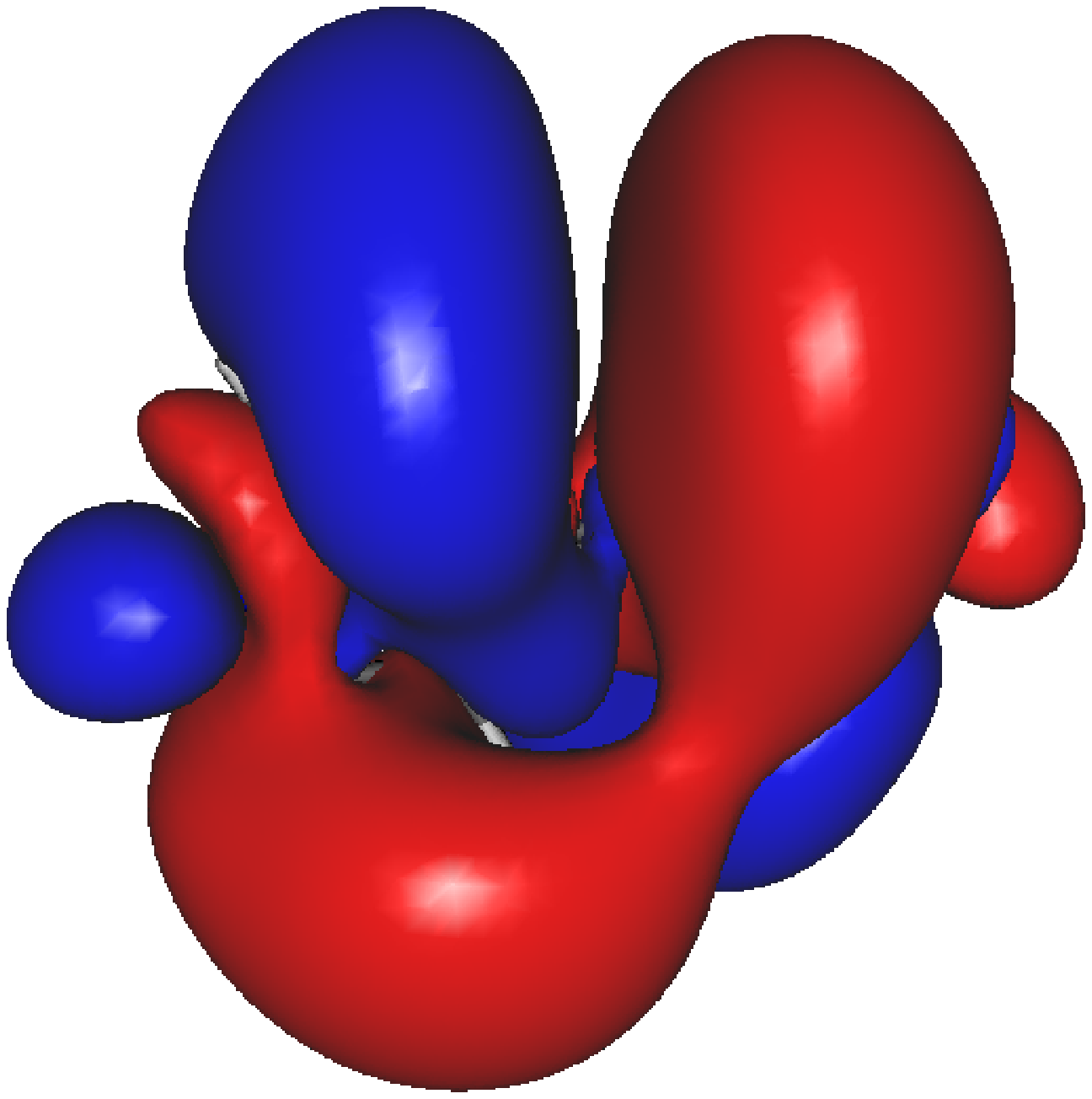}\hspace*{0.ex}}
$ $\\[-1.4ex]
\caption{The almost singly occupied natural orbital corresponding to the anion's extra electron 
of the 44BPY$^{\bullet -}$ anion (``LUMO'')
obtained via EA-EOM-CCSD/aug-cc-pVDZ calculations is delocalized over the whole molecule.}
\label{fig:lumo-44bpy}
\end{figure}

Eq.~(\ref{eq-phi-i}) cannot be directly applied to a real molecular junction.
Contrary to the usual claim \cite{Neaton:06,Quek:07,Venkataraman:12a}, 
for cases relevant for molecular electronics \cite{Baldea:2013b,Baldea:2014c}, 
realistic LUMO's are not point-like but rather extended over the entire molecule. 
LUMO spatial distributions $\rho^{LUMO}$ of the two molecules considered in this study
are shown in Figures \ref{fig:lumo-bdcn} and \ref{fig:lumo-44bpy}. 
Because spatial densities of 
Kohn-Sham LUMO's are completely unphysical
and Hartree-Fock LUMO's may represent a too crude approximation, 
like in refs.~\cite{Klaiman:13,Baldea:2014c}, 
we have calculated the natural orbital expansion of the 
corresponding reduced density matrices at the EOM-CCSD level.
For the extra electron, 
we obtained that a single natural orbital almost entirely exhausts the natural 
orbital expansion; for BDCN and 44BPY, the weights are 98.1\% and 97.7\%, respectively. 
So, this method is indeed best suited to describe the spatial distribution of the extra electron 
in molecules with n-type (LUMO-mediated) conduction.

%%%%%%%%%%%%%%%%%%%%%%%%%%%%%%%%%%%%%%%%%%%%%%%%%%%%%%%%%%%%%%%%%%%%%%%%%%%%%%%% 
%%%%%%%%%%%%%%%%%%%%%%%%%%%%%%%%%%%%%%%%%%%%%%%%%%%%%%%%%%%%%%%%%%%%%%%%%%%%%%%%

%%%%%%%%%%%%%%%%%%%%%%%%%%%%%%%%%%%%%%%%%%%%%%%%%%%%%%%%%%%%%%%%%%%%%%%%%%%%%%%% 
%%%%%%%%%%%%%%%%%%%%%%%%%%%%%%%%%%%%%%%%%%%%%%%%%%%%%%%%%%%%%%%%%%%%%%%%%%%%%%%%
As visible in Figures \ref{fig:lumo-bdcn} and \ref{fig:lumo-44bpy},
rather than being strongly peaked close to the center
(which would have justified to assume a point-like LUMO), 
the natural orbital densities $\rho^{LUMO}(\mathbf{r})$ of 
the extra electron is found to be spread over the whole molecules. 
Therefore, the LUMO energy shift driven by image charges should be calculated by appropriately
weighting Eq.~(\ref{eq-phi-i})
%%%%%%%%%%%%%%%%%%%%%%%%%%%%%%%%%%%%%%%%%%%%%%%%%%%%%%%%%%%%%%%%%%%%%%%%%%%%%%
\begin{equation}
\label{eq-vIm}
\Phi_{im} = \int_{z_s}^{z_t} d z \,  \rho_{1D}^{LUMO}(z)\, \phi_{im} (z) ,
\end{equation}
%%%%%%%%%%%%%%%%%%%%%%%%%%%%%%%%%%%%%%%%%%%%%%%%%%%%%%%%%%%%%%%%%%%%%%%%%%%%%%
where $\rho_{1D}^{LUMO}(z) = \int d x \, d  y \, \rho^{LUMO}(\mathbf{r})$ is the LUMO density
along the molecular axis $z$. 

\begin{figure}[h!]
$ $\\[0ex]
\centerline{\hspace*{0.ex}\includegraphics[width=0.4\textwidth,angle=0]{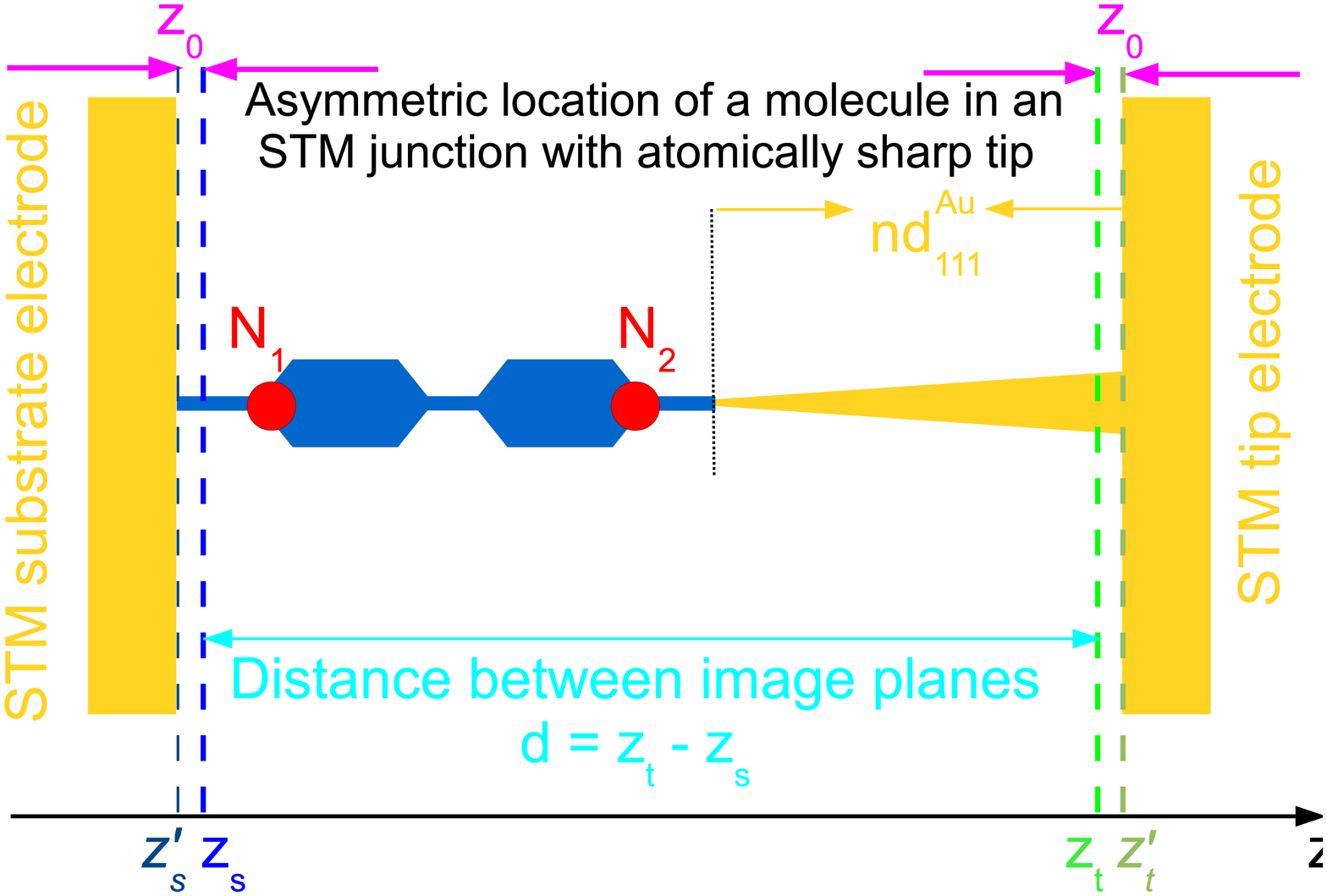}\hspace*{0.ex}
\includegraphics[width=0.4\textwidth,angle=0]{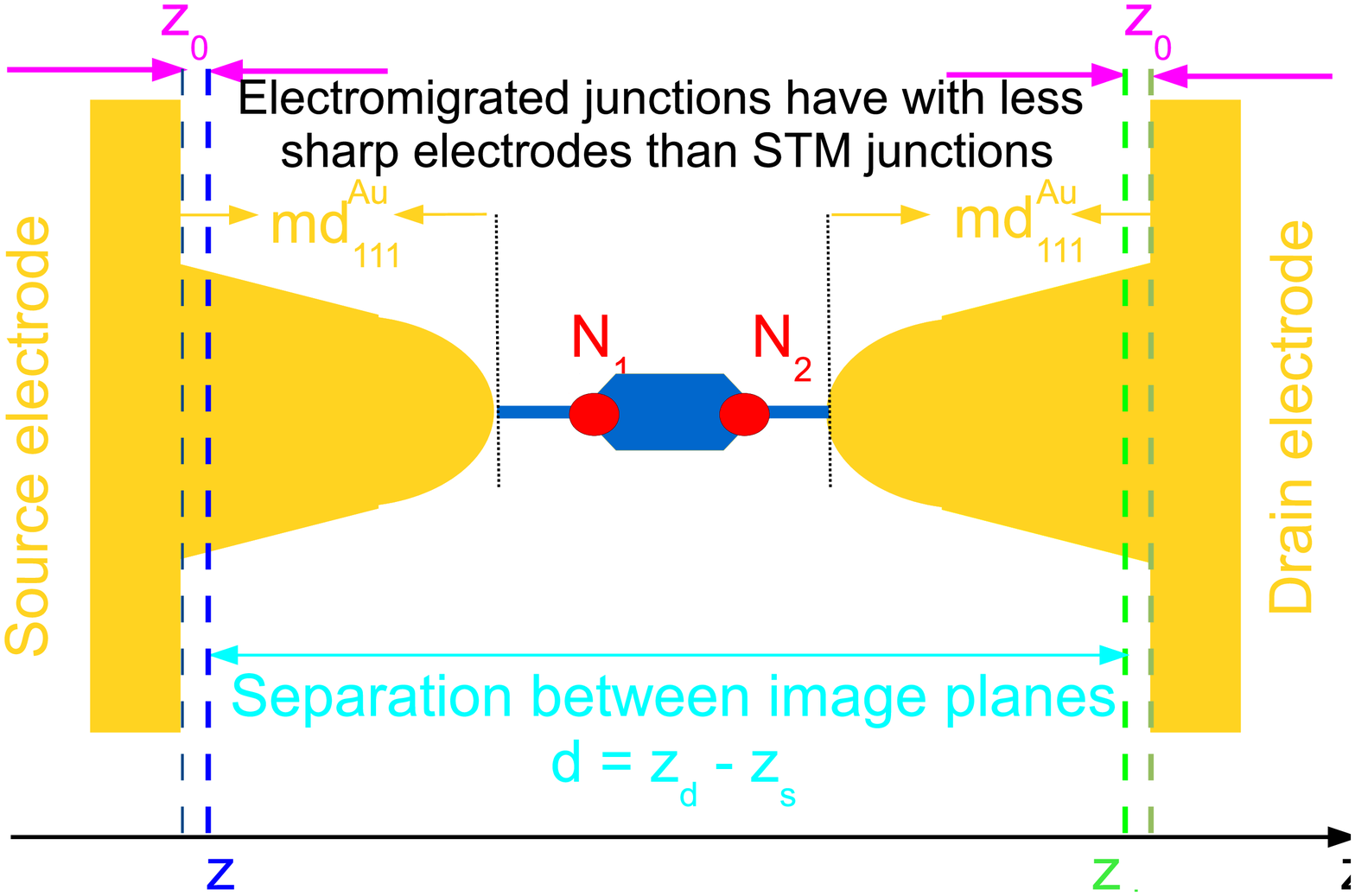}\hspace*{0.ex}
}
$ $\\[-1.4ex]
\caption{Cartoons illustrating the basic features of an asymmetric STM setup with a sharp STM-tip 
\cite{Venkataraman:12a} (left panel) and the less sharp electrodes of a rather symmetric 
electromigrated junction \cite{Reed:09} (right panel).}
\label{fig:cartoons}
\end{figure}

For properly estimate the image-driven shifts $\Phi_{im, 1} \equiv \Phi_{im, BDCN}$ and
$\Phi_{im, 2} \equiv \Phi_{im, 44BPY}$ via Eq.~(\ref{eq-vIm}), 
attention should be paid to the difference between the 
experimental setups employed in refs.~\cite{Reed:09} and \cite{Venkataraman:12a}, respectively.
This difference is illustrated by the two cartoons in the left and right panels of 
Figure \ref{fig:cartoons}. 
%%%%%%%%%%%%%%%%%%%%%%%%%%%%%%%%%%%%%%%%%%%%%%%%%%%%%%%%%%%%%%%%%%%%%%%%%%%%%%%% 
%%%%%%%%%%%%%%%%%%%%%%%%%%%%%%%%%%%%%%%%%%%%%%%%%%%%%%%%%%%%%%%%%%%%%%%%%%%%%%%%

In an asymmetric STM-setup like that used in the measurements
for 44BPY-based junctions considered here \cite{Venkataraman:12a}, the usual assumption 
\cite{Neaton:06,Quek:07,Venkataraman:12a} of an infinite 
plate in the immediate vicinity of the (nitrogen) atom at one molecular end 
is justified only for one electrode (STM substrate).
As shown recently \cite{Baldea:2013b}, to model an atomically sharp (\emph{e.g.}, pyramidal) STM-tip
with a height $n d_{111}^{Au}$, 
one can consider an infinite planar electrode displaced from the tip appex by an effective 
number of $n$ 
Au(111) layers. That is (for atom notation see Figure \ref{fig:cartoons}, 
$z_s = z_{N_1} - d_{Au-N} + z_0$ 
and $z_t = z_{N_2} + d_{Au-N} - z_0 + n d_{111}^{Au}$  
($d_{111}^{Au} \simeq 2.354$\,{\AA}, $d_{Au-N} \simeq 2.336$) \cite{Baldea:2013b}.
An estimate
%%%%%%%%%%%%%%%%%%%%%%%%%%%%%%%%%%%%%%%%%%%%%%%%%%%%%%%%%%%%%%%%%%%%%%%%%%%%%%%%%%%%%%%%%%
\begin{equation}
\label{eq-Phi-2}
\left . \Phi_{im, 2}\right\vert_{n \approx 3} \approx -1.39\mbox{\,eV}
\end{equation}
%%%%%%%%%%%%%%%%%%%%%%%%%%%%%%%%%%%%%%%%%%%%%%%%%%%%%%%%%%%%%%%%%%%%%%%%%%%%%%%%%%%%%%%%%%
is obtained by using $n \approx 3$, a value that turned out to be in excellent agreement with 
the experimental data on 44BPY-based junctions analyzed in ref.~\cite{Baldea:2013b}. 

In the same spirit, to model the (basically symmetric) experimental setup of the electromigrated BDCN-based 
junctions of ref.~\cite{Reed:09}, we will use $z_s = z_{N_1} - d_{Au-N} + z_0 - m d_{111}^{Au}$ and
$z_s = z_{N_2} + d_{Au-N} - z_0 + m d_{111}^{Au}$, which amounts to consider image planes 
displaced by an effective number $m$ of Au(111) layers. 
Taking a value of $m$ smaller than $n$ is in accord to the fact that the two gold electrodes 
in an electromigrated setup \cite{Reed:09} are not so sharp as an STM tip \cite{Venkataraman:12a}.
Therefore, to get a simple estimate we will assume $m \simeq n/2$, 
which yields via Eq.~(\ref{eq-vIm})
%%%%%%%%%%%%%%%%%%%%%%%%%%%%%%%%%%%%%%%%%%%%%%%%%%%%%%%%%%%%%%%%%%%%%%%%%%%%%%%%%%%%%%%%%%
\begin{equation}
\label{eq-Phi-1}
\left . \Phi_{im, 1} \right \vert_{m \approx 1.5} \approx -0.68\mbox{\,eV} .
\end{equation}
%%%%%%%%%%%%%%%%%%%%%%%%%%%%%%%%%%%%%%%%%%%%%%%%%%%%%%%%%%%%%%%%%%%%%%%%%%%%%%%%%%%%%%%%%%
%%%%%%%%%%%%%%%%%%%%%%%%%%%%%%%%%%%%%%%%%%%%%%%%%%%%%%%%%%%%%%%%%%%%%%%%%%%%%%%%
\subsection{Validation of the disentangling scheme}
\label{sec:validate}
\begin{figure}[h!]
$ $\\[5ex]
\centerline{\hspace*{0.ex}\includegraphics[width=0.5\textwidth,angle=0]{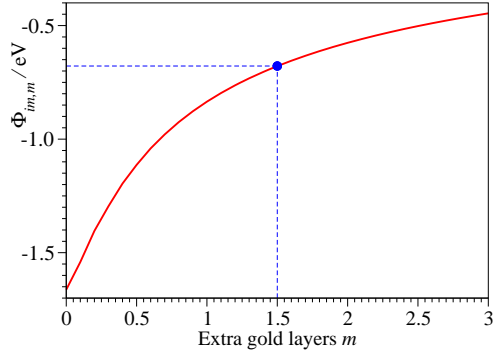}\hspace*{0.ex}}
$ $\\[-1.4ex]
\caption{Dependence on the effective number ($m$) of extra gold layers of the image-driven LUMO energy shift 
$\Phi_{im, m}$ in an electromigrated junction, 
as schematically presented in the right panel of Figure \ref{fig:cartoons}. See the main text for details.}
\label{fig:vIm-bdcn}
\end{figure}

%%%%%%%%%%%%%%%%%%%%%%%%%%%%%%%%%%%%%%%%%%%%%%%%%%%%%%%%%%%%%%%%%%%%%%%%%%%%%%%%%%%%%%%%%
With the numerical values given by Eqs.~(\ref{eq-e0-1}), (\ref{eq-e0-2}), (\ref{eq-Phi-1}), and
(\ref{eq-Phi-2}) and the first line of Table \ref{table}, the following values of the LHS and RHS 
of Eq.~(\ref{eq-key}) are obtained 
%%%%%%%%%%%%%%%%%%%%%%%%%%%%%%%%%%%%%%%%%%%%%%%%%%%%%%%%%%%%%%%%%%%%%%%%%%%%%%%%%%%%%%%%%%
\begin{eqnarray}
\label{eq-LHS}
\varepsilon_{0, 2} - \varepsilon_{0, 1} & = & 0.07 \pm 0.12 \mbox{\,eV} , \\
\label{eq-RHS}
EA_{1} - EA_{2} + \Phi_{im, 2} - \Phi_{im, 1} & \approx & -0.02 \mbox{\,eV} .
\end{eqnarray}
%%%%%%%%%%%%%%%%%%%%%%%%%%%%%%%%%%%%%%%%%%%%%%%%%%%%%%%%%%%%%%%%%%%%%%%%%%%%%%%%%%%%%%%%%%
So, the values of Eqs.~(\ref{eq-LHS}) and (\ref{eq-RHS}) are
in accord with Eq.~(\ref{eq-key}) within errors. Concerning the estimates of 
Eqs.~(\ref{eq-Phi-2}) and (\ref{eq-Phi-1}),
we note that they are not substantially affected by the values chosen above for $n$ and $m$.
Since the robustness with respect to reasonable changes in $n$ has been analyzed 
in ref.~\cite{Baldea:2013b}, we only present here the $m$-dependence
(see Figure \ref{fig:vIm-bdcn}).
%%%%%%%%%%%%%%%%%%%%%%%%%%%%%%%%%%%%%%%%%%%%%%%%%%%%%%%%%%%%%%%%%%%%%%%%%%%%%%%% 
%%%%%%%%%%%%%%%%%%%%%%%%%%%%%%%%%%%%%%%%%%%%%%%%%%%%%%%%%%%%%%%%%%%%%%%%%%%%%%%%
Based on these results, we conclude that 
the above estimates for $\Phi_{im, 1}$ and $\Phi_{im, 2}$ are accurate within $\sim 0.1$\,eV,
which is consistent to experimental inaccuracies expressed in Eq.~(\ref{eq-LHS}). 

To end this section, we note that, like for 44BPY-based junctions 
(see Figure 4(b) of ref.~\cite{Baldea:2013b}), 
gold atoms linked to 
a BDCN molecule do not substantially affect the LUMO spatial distribution.
This aspect, which is visible in Figure \ref{fig:vIm-bdcn}, is relevant: 
it demonstrates that, for the molecules under consideration, 
corrections due to image charges are not
dramatically affected by the cutoff procedure close to electrodes.
%%%%%%%%%%%%%%%%%%%%%%%%%%%%%%%%%%%%%%%%%%%%%%%%%%%%%%%%%%%%%%%%%%%%%%%%%%%%%%%%
\section{Conclusion}
\label{sec:conclusion}
%%%%%%%%%%%%%%%%%%%%%%%%%%%%%%%%%%%%%%%%%%%%%%%%%%%%%%%%%%%%%%%%%%%%%%%%%%%%%%%%
The energetic alignment $\varepsilon_0$ relative to electrodes' Fermi energy of 
the dominant orbital represents a key quantity that controls the charge transport 
by tunneling in molecular junctions. 
% Proposing and validating disentangling schemes 
Disentangling $\varepsilon_0$ in 
contributions with clear physical origin may be important
not only for fundamental nanoscience but also for designing electronic nanodevices.

Methodologically, to validate such disentangling schemes, it is preferable to 
compare two molecular junctions that basically differ in a \emph{single} respect.
The disentangling scheme analyzed recently \cite{Baldea:2013b}, where we have considered junctions 
based on the same molecule (44BPY) but placed in different environments 
(solvent \cite{Tao:03} versus ambient conditions \cite{Venkataraman:12a}), 
was a first step in this direction.
As a further step in the same vein, in this paper we have considered a disentangling scheme for 
nanojunctions, wherein the two molecules considered (BDCN and 44BPY) are different but their
chemical linkage to electrodes is similar (nitrogen-gold affinity). 

\ib{
We believe that the validation of the presently proposed disentangling scheme against available experimental 
data \cite{Reed:09,Venkataraman:12a} is noteworthy. Still, considering more transport data for 
molecular junctions exhibiting n-type conduction to generalize the proposed method beyond the two aforementioned cases
is highly desirable. $I-V$ curves for extended bias ranges well beyond the Ohmic regime ($\vert V\vert \agt V_t$), 
allowing to determine the transition voltage $V_t$ and thence the MO energy offset $\varepsilon_0$ 
[\emph{cf.~}Eq.~(\ref{eq-vt})], supplemented 
by thermopower data \cite{Venkataraman:12a} or employing electrodes with different work functions \cite{Frisbie:11}
as evidence for a (LUMO-mediated, n-)type of conduction, would be best suited for this purpose. 
Oligophenylenes with isocyanide linkages NC-(C$_6$H$_4$)$_n$-CN,
\emph{i.e.}~series with several ($n$) phenylene rings instead of the single ($n=1$) ring of the 
BDCN$\equiv$NC-C$_6$H$_4$-CN considered above and in ref.~\cite{Reed:09}, 
may represent good candidates for such investigations.
Reliable quantum chemical methods like those used here or elsewhere \cite{Baldea:2014c} can still be applied 
for molecular species with up to $n=3-4$ rings. 
Unfortunately, we were unable to find such experimental transport data in existing publications, 
which are very often restricted to the
Ohmic conductance. Still, we hope that the present theoretical study 
will encourage accompanying experimental (and further theoretical) efforts to validate similar disentangling scheme 
that could certainly contribute to a better microscopical understanding of the nanotransport.
}

We end with the following technical remark. To validate the disentangling scheme proposed 
in ref.~\cite{Baldea:2013b} we have resorted to $\Delta$-DFT calculations. 
As compared to more elaborate quantum chemical methods, the $\Delta$-DFT method 
is computationally considerably less demanding. 
As revealed by the comparison between the first and the last line
in Table \ref{table} and also discussed elsewhere \cite{Baldea:2014c}, $\Delta$-DFT-based estimates 
for $\varepsilon_0$ of a \emph{given} molecular junction may not be satisfactory. 
Still, \emph{differences} $\varepsilon_{0,1} - \varepsilon_{0,2}$ between relevant MO offsets 
$\varepsilon_{0,1}$ and $\varepsilon_{0,2}$ characterizing different (but not too different) molecular 
junctions estimated within $\Delta$-DFT can be of an accuracy comparable to those 
based on the computationally very costly EOM-CCSD, which represents the 
state-of-the-art of quantum chemistry of medium size molecules.
This is also an important aspect, as it allows to understand \emph{differences} between 
properties of various nanojunctions by resorting to lower cost computational approaches.
%%%%%%%%%%%%%%%%%%%%%%%%%%%%%%%%%%%%%%%%%%%%%%%%%%%%%%%%%%%%%%%%%%%%%%%%%%%%%%%%
\section*{Acknowledgments}
Financial support for this work provided by the 
Deutsche For\-schungs\-ge\-mein\-schaft 
(grant BA 1799/2-1) is gratefully acknowledged.
%%%%%%%%%%%%%%%%%%%%%%%%%%%%%%%%%%%%%%%%%%%%%%%%%%%%%%%%%%%%%%%%%%%%%%%%%%%%%%%%

\end{document}